\documentclass[twocolumn,aps,prl,superscriptaddress]{revtex4}
\usepackage{graphicx}

\begin{document}
\pagestyle{empty}

\title{How do liquids confined at the nanoscale influence adhesion?}

\author{C. Yang}
\affiliation{IFF, FZ-J\"ulich, 52425 J\"ulich, Germany}
\author{U. Tartaglino}
\affiliation{IFF, FZ-J\"ulich, 52425 J\"ulich, Germany}
\affiliation{DEMOCRITOS National Simulation Center, Via Beirut 2, 34014 Trieste, Italy}
\author{B.N.J. Persson}
\affiliation{IFF, FZ-J\"ulich, 52425 J\"ulich, Germany}

\begin{abstract}

Liquids play an important role in adhesion and sliding friction. 
They behave as lubricants in human bodies especially in the joints.
However, in many biological attachment systems they acts like
adhesives, e.g.  facilitating insects to move on ceilings or 
vertical walls. Here we use molecular dynamics to study how
liquids confined at the nanoscale influence the adhesion between 
solid bodies with smooth and rough surfaces. We show that
a monolayer of liquid may strongly affect the adhesion.
\vspace{1em}\\
{\it Reference:} J. Phys.: Condens. Matter {\bf 18}, 11521 (2006)\\
{\it DOI:} 10.1088/0953-8984/18/50/008 \\
{\it Preprint} Arxiv:cond-mat/0608098

\end{abstract}

\maketitle


Knowing the behavior of liquids confined to small volumes between contacting
surfaces is essential for the understanding a vast array of common problems
in science, such as biological interactions\cite{Gorb,Bo_JCP,Bo_squeeze_out}, 
crack propagation\cite{NATO},
molecular tribology and adhesion\cite{Bo_squeeze_out,Bo_book,Is,Arzt,Science_liquid}.
Ultrathin confined films can have two completely 
opposite effects.
On one hand, the confined liquid may behave like a lubricant to reduce adhesion
and friction. For instance, when bones meet at a joint, they need a liquid
in-between to prevent scraping against each other. The liquid is called
a synovial liquid \cite{Synovial}, 
which is made mainly of water and water-based long chain polymers. 
Many internal organs in humans and other animals are separated by thin 
lubricant films. Examples include the cerebrospinal liquid
in the brain \cite{Cerebro}, the pleural liquid in lungs \cite{Pleural} 
and water based liquid in eyes.
On the other hand, in some applications liquids behave like adhesives, 
e.g. in some biological attachment systems. Thus, for example, 
insects can secret a liquid to enhance adhesion in order to walk on 
ceilings or vertical walls \cite{Gorb}.
Geckos can also walk on the ceilings due to their hierarchical hair 
structure which makes their adhesion pads elastically 
soft on all relevant length scales\cite{Autumn}. 
Recently scientists found that 
water monolayer-films, which always occur on
hydrophilic (water-loving) surfaces in the normal atmosphere, enhance 
Gecko adhesion \cite{Huber,thesis} to hydrophilic surfaces. 

The influence of thin liquid layers on the effective adhesion
between solids is well-known to most of us from our every-day experience, e.g., when separating
two (slightly wet) microscopy cover glass surfaces, 
or two gauge blocks (steel blocks with very smooth
surfaces). At the micro or nano-scale the increased adhesion resulting from the formation of
water capillary bridges is one of the most severe problem in the production of 
micro-electro-mechanical systems (MEMS)\cite{Zhao,Hydration,MEMS}.  
In general, wetting liquids tend to increase the adhesion between solids with surface roughness,
but at present it is not known exactly for which liquid coverage 
the pull-off force
is maximal\cite{Is,JCP2004}. There are also many unsolved questions related to 
the influence of liquids on adhesion in biological systems,
e.g., how is it possible for a Gecko to move on a vertical 
stone wall during raining, and how do cells (in liquids) adhere to solids walls?

In this study we show
that small amounts of confined liquid between smooth
and randomly rough surfaces can
influence adhesion. 
Using liquids that interact weakly and strongly 
with the walls (hydrophobic \cite{Drop} and hydrophilic walls),
we demonstrate that sometimes the liquid acts like a lubricant,
as in some biological systems, such as in the eyes, 
and sometimes it behaves like a glue, as the insects' secretions
when they walk on ceilings or vertical walls.

We have used molecular dynamics calculation to study the influence of
liquids confined at the nanoscale on adhesion. We have simulated 
an elastic block approaching a hard substrate covered with
octane molecules. The atoms in the bottom layer of the block form
a simple square lattice with lattice constant $a=2.6$ \AA. 
The lateral dimensions $L_{x}=N_{x}a$ and $L_{y}=N_{y}a$. For the block,
$N_{x}=N_{y}=48$. Periodic boundary conditions are applied in the
$xy$ plane. The lateral size of the block is equal to that of substrate, 
but we used different a lattice constant $b\approx a/\phi$, 
where $\phi=(1+\sqrt5)/2$ is the golden mean,
in order to avoid the pinning due to a perfect match of the crystal structures.
The thickness of the block, 142 \AA, is comparable to the lateral length
of the system.
A recently developed Multiscale Molecular Dynamics description
\cite{Yang} let us properly describe the elastic response of the block
without having to simulate too many atoms in the bulk. The Young modulus
of the block is $E=100$ GPa and the Poisson ratio is $\nu=0.3$.
We consider both atomically flat and rough substrates. 
The profile of the rough substrate is self affine fractal with root-mean-square
(rms) amplitude 3 \AA\ and fractal dimension $D_{\mbox{f}}=2.2$;
it is prepared following the recipe in Refs.~\cite{Yang,P3}.

The liquid between the walls is octane C$_8$H$_{18}$.
A monolayer of liquid on the flat substrate is composed of $409$ octane
molecules.
The simulations are carried out at room temperature, $T=300$ K,
well above the melting point of octane $T_{m}=216$ K.
The octane molecules are treated with the United Atom model: every
molecule comprises 8 particles corresponding to the groups CH$_2$ and
CH$_3$. The Optimized Potential for Liquid Simulation (OPLS)
\cite{jorgensen1984x1,dysthe2000x1} is employed.
The atoms of the two walls interact through the Lennard-Jones potential 
\cite{Yang}, with parameters $\sigma=3.28$ \AA\ and $\epsilon=40$ meV.
The interaction between the walls' atoms and the groups CH$_2$ and
CH$_3$ of the octane molecules are also given by the Lennard-Jones potential
with $\sigma=3.28$ \AA, but two different interaction energies are
considered: the \emph{strong adsorbate interaction} with $\epsilon=40$ meV,
and the \emph{weak adsorbate interaction} with $\epsilon=5$ meV.

\begin{figure}
  \includegraphics[width=0.35\textwidth]{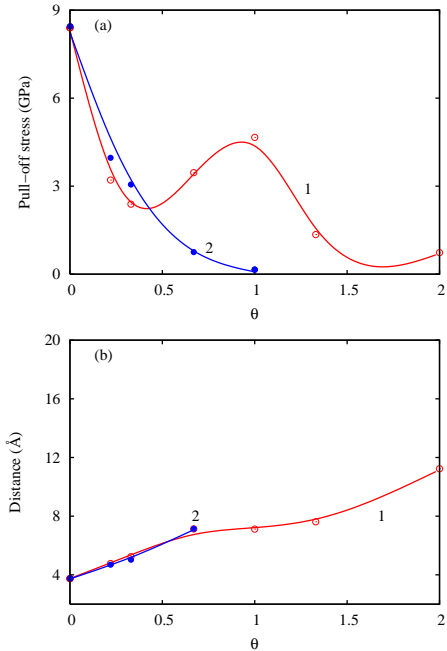} 
  \caption{ \label{adhesion_flat}
     Pull-off on a flat substrate for strong adsorbate interaction
     (curve {\bf 1}) and for weak adsorbate interaction (curve {\bf 2}).
     (a): the maximum pull-off stress
     versus the liquid coverage $\theta$.
     (b): the separation between the walls where the stress is maximum
     versus $\theta$.
     Continuous lines: fit with smoothing cubic splines.
  }
\end{figure}

In the simulations we first let the octane deposit on the substrate and
thermalise, then we bring the elastic block in contact with the substrate,
and finally we pull up the block slowly until the two surfaces separate.
The overall preparation of the system before retraction requires the
simulation of 4 to 6 nanoseconds. Approach and retraction are
realized by imposing a fixed speed $v=1$ m/s to the outer surface
of the block, while the wall in contact with the substrate and
the bulk of the block are free to deform elastically.
The pulling speed is small enough not to alter the pull-off process.
Indeed we performed simulations (not shown) to detect when the
pulling speed becomes relevant: we observed that the pull-off force
changes significantly only for speeds above 50 m/s.
We speculate that for larger amounts of liquids and for chains longer than
octane (more cross-linked) the influence of speed can appear 
at smaller velocities.
In the process of detachment the interaction between the
lubricated walls is attractive, and the adhesive force reaches a maximum
at a given distance, which depends on the amount of liquid,
and then it decreases.
Fig.~\ref{adhesion_flat}(a) shows the maximum pulling force per unit area
(pull-off stress), as a function of liquid
coverage $\theta$, for the case of two flat walls. The corresponding
distance between the walls when the pulling force is maximum is plotted
in figure \ref{adhesion_flat}(b).

\begin{figure}
  \includegraphics[width=0.3\textwidth]{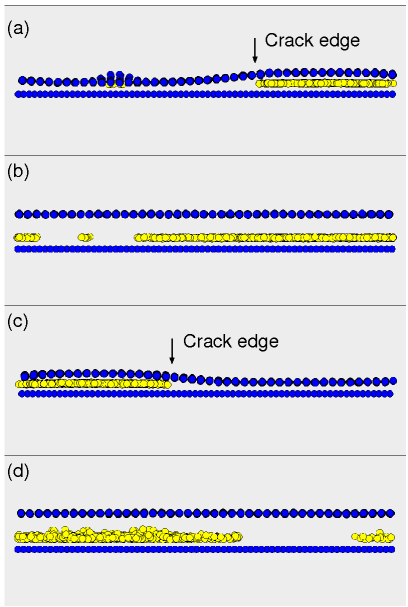} 
  \caption{ \label{s3_flat_jump_snapshots}
     Side view of the system with coverage $\theta\approx 1/4$ during
     the pull-off. The light balls represent the CH$_2$ and CH$_3$
     groups of the octane molecules. The dark balls are the atoms of the
     walls' surfaces.
     (a) Strong adsorbate interaction between the liquid and the walls;
     the separation happens by crack propagation.
     (b) After the detachment the block returns to its undeformed state.
     (c) and (d) during and after the detachment for the case of weak
     adsorbate interaction. The behaviour is the same as in the previous
     case.
  }
\end{figure}

We note that the direct interaction between the block and the substrate
is negligible when the surfaces are separated by one adsorbate monolayer. 
The pull-off stress is maximum for clean and smooth surfaces since
there is strong interaction between the wall atoms
(Lennard-Jones parameter $\epsilon=40$ meV).
The pull-off stress rapidly decreases 
with increasing amount of liquid between the surfaces.
For $\theta<1$, the boundary line of an island of lubricant
molecules can be considered as a crack edge, as shown in
Fig.~\ref{s3_flat_jump_snapshots}. During pull-off high stress
concentration occurs at the crack tip, 
and the interfacial crack propagation starts at the crack edge,
resulting in the strong decrease in the pull-off force observed in the
simulations for $\theta < 1$.

Fig.~\ref{s3_flat_jump_snapshots} shows the side view of the
system before and after detachment when the coverage is $\theta=1/4$.
For the strong interaction between liquid and walls [(a) and (b)]
not all molecules are trapped in one island.
For the  weak adsorbate interaction [(c) and (d)]
the liquid molecules experience smaller lateral energy barriers, and 
for $\theta < 1$ they arrange themselves into a single large
island, and thus less elastic energy is stored at the block-substrate
interface in this case, resulting, at low liquid coverage, 
in a (slightly) stronger adhesion (or pull-off force) 
for the weak adsorbate interaction as compared to the case of
strong interaction.
The morphology of the adsorbate depends on the preparation of the system.
In an infinitely large system the liquid will always form finite clusters,
whose size and density are determined by the competition between
deposition and diffusion \cite{Jensen}.
As a general rule the stronger interaction
will give rise to smaller islands, so that for the same preparation
the stronger adsorbate interaction will lead to a slightly lower adhesion,
as shown by our simulations.
Notice that after the detachment of the two walls all the liquid sticks
to the substrate: indeed its adhesion energy to the substrate is
larger because of the larger substrate 
atom density compared to that of the upper
block.

For the weak adsorbate interaction the pull-off stress 
decreases monotonically with increasing C$_8$H$_{18}$
coverage. On the other hand the strong adhesive interaction causes an
increase of the pull-off stress up to a maximum for the coverage
$\theta=1$ corresponding to a complete monolayer.
The physical reason for this phenomenon is that at $\theta=1$
the liquid completely covers the atomically smooth surface and no
crack-like defects occur at the interface; moreover the stored elastic
energy at the interface is minimal. Thus the adhesion reaches a local
maximum.

\begin{figure}
  \includegraphics[width=0.35\textwidth]{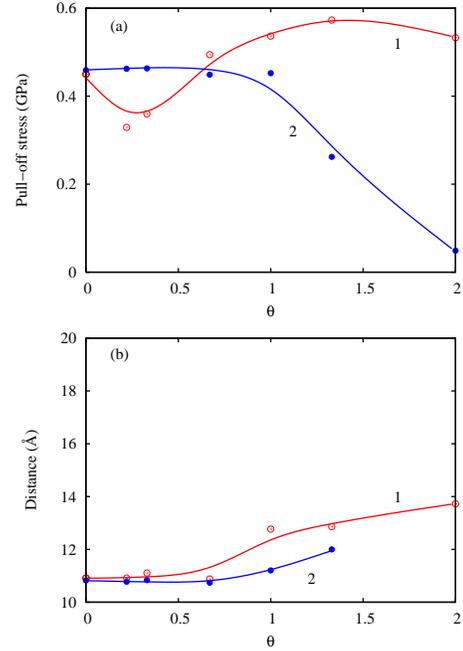} 
  \caption{ \label{adhesion_rough}
     Pull-off on a rough substrate for strong adsorbate interaction
     (curve {\bf 1}) and for weak adsorbate interaction (curve {\bf 2}).
     (a): the maximum pull-off stress
     versus the liquid coverage $\theta$.
     (b): the walls distance where the stress is maximum
     versus $\theta$.
     Continuous lines: fit with smoothing cubic splines.
  }
\end{figure}

\begin{figure}
  \includegraphics[width=0.3\textwidth]{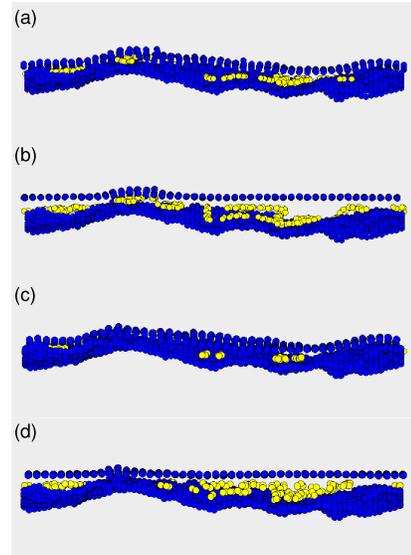} 
  \caption{ \label{s3_rough_trapped_molecules}
           Side view of the contact for coverage $\theta=1/4$ on
           a rough substrate.
           The pictures show only the atoms in a 29 \AA\ thick slice
           including the highest asperity.   
           (a) Strong adsorbate interaction; initial state.
           (b) Strong adsorbate interaction; maximum pull-off force.
           (c) Weak adsorbate interaction; initial state.
           (d) Weak adsorbate interaction; maximum pull-off force.}
\end{figure}

\begin{figure}
  \includegraphics[width=0.3\textwidth]{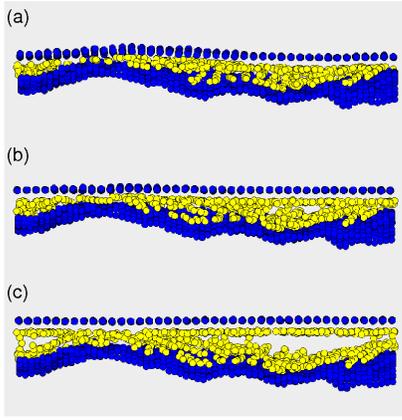} 
  \caption{ \label{r_b2_ep40_snap}
           Side view of some atoms for the strong interacting adsorbate
           at coverage $\theta=1.3$ on a rough substrate, including the
           highest asperity.
           The system evolves in time from (a) to (b) and then to (c)
           during the pulling phase.
           }
\end{figure}

For the rough (self-affine fractal) substrate, 
the pull-off stress versus $\theta$ is showed in Fig.~\ref{adhesion_rough}.
In this case the 
monolayer coverage corresponds to $\theta \approx 1.3$
i.e., the roughness increases the substrate surface area by about $30\%$.
For the strong adsorbate interaction, 
the adhesion decreases with increasing liquid
coverage up to $\theta \approx 0.2$ due to molecules trapped in the asperity
contact regions,
which effectively enhances the substrate surface roughness. 
For $\theta > 0.2$ the pull-off force increases due 
to the formation of capillary bridges (see Fig.~\ref{r_b2_ep40_snap} (c)). 
For the weak adsorbate interaction, 
due to the strong wall-wall
interaction and weak liquid-wall interaction,
the liquid molecules are squeezed away from the
asperity contact regions into the valleys 
[see Fig.~\ref{s3_rough_trapped_molecules} (c),(d) and  
Fig.~\ref{r_b2_topview_snap}], which results in
nearly constant pull-off force for $\theta < 1$ 
(see curve {\bf 2} in Fig.~\ref{adhesion_rough}).
Due to the hydrophobic interaction between the liquid and the blocks,
no capillary bridge forms, and as $\theta$ increases beyond $1$ the 
pull-off force decreases towards zero.

\begin{figure}
  \includegraphics[width=0.3\textwidth]{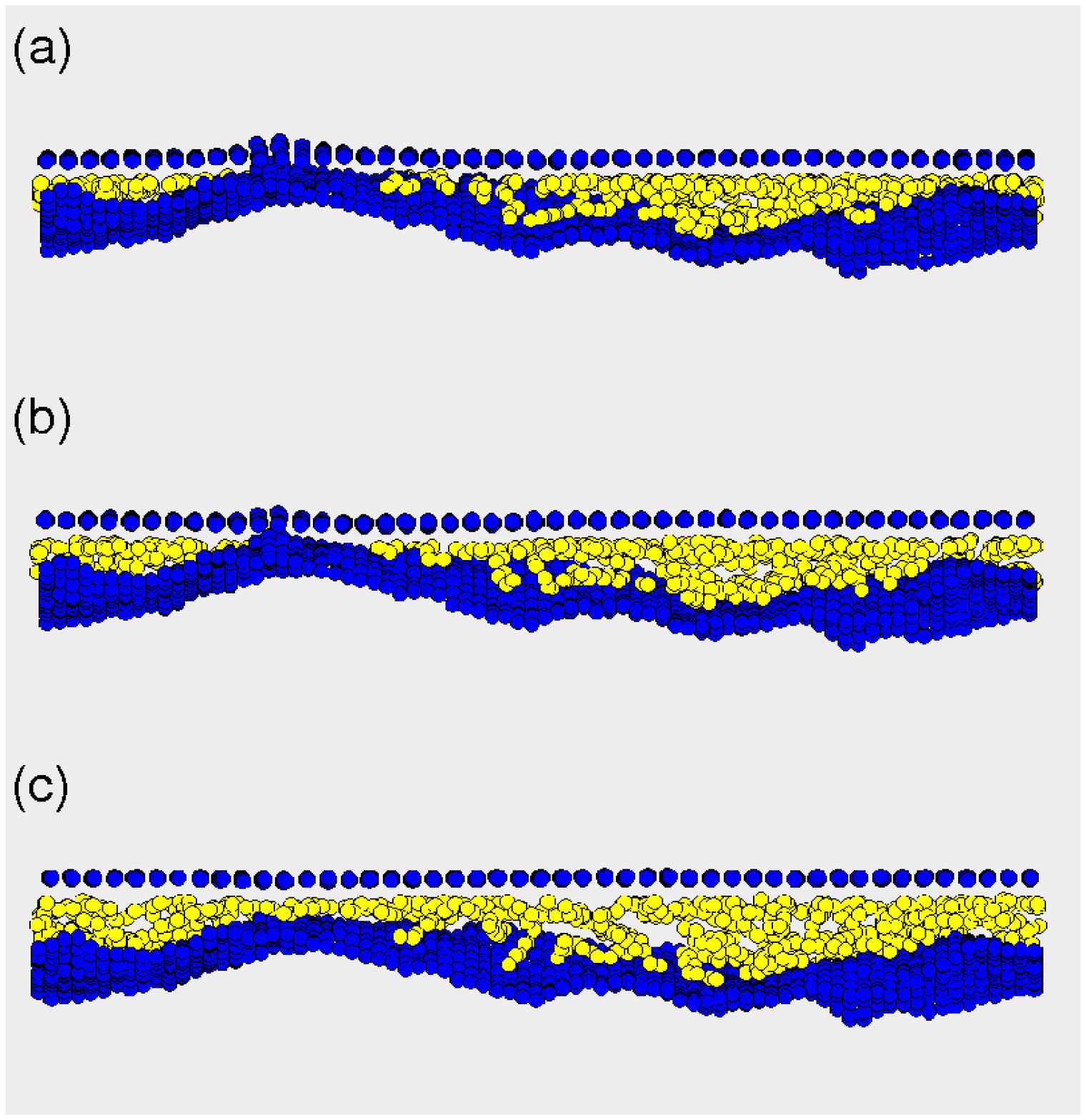} 
  \caption{ \label{r_b2_ep5_snap}
           Side view of some atoms for the weak interacting adsorbate
           at coverage $\theta=1.3$ on a rough substrate, including the
           highest asperity.
           The system evolves in time from (a) to (b) and then to (c)
           during the pulling phase.
  }
\end{figure}

\begin{figure}
  \includegraphics[width=0.5\textwidth]{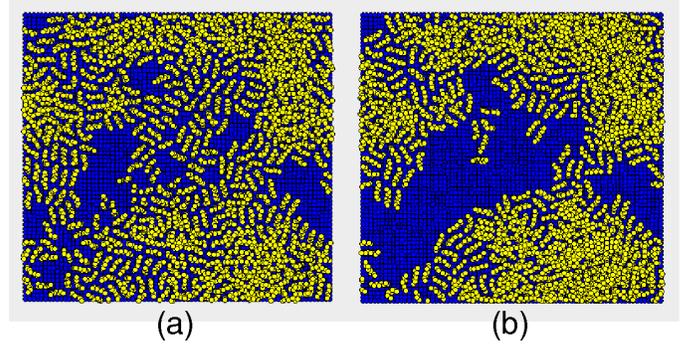} 
  \caption{ \label{r_b2_topview_snap}
          Top view snapshots of the liquid on a rough substrate
          for $\theta \approx 1.3$,
          with (a) strong interaction and (b) weak interaction between liquid
          and walls. These images correspond to the side views shown in
          Fig.~\protect\ref{r_b2_ep40_snap}(a) and
          Fig.~\protect\ref{r_b2_ep5_snap}(a) respectively.   
  }
\end{figure}

For the strong adsorbate interaction 
the lateral corrugation of the molecule-substrate interaction potential 
is so high that 
liquid molecules are trapped in the asperity contact regions.
This is illustrated in Fig.~\ref{s3_rough_trapped_molecules} (a) 
and (b) (side-view snapshots). The trapped molecules increase the elastic energy
stored at the interface and reduce the effective wall-wall binding energy.
Thus, for $\theta < 0.67$ 
the pull-off stress for the strong adsorbate interaction
is smaller than that for the weak interaction, as it is shown
in Fig.~\ref{adhesion_rough}(a).
For the rough substrate with strong adsorbate interaction 
the pull-off stress reaches 
a maximum at $\theta \approx 1.3$ (see curve {\bf 1} in Fig.~\ref{adhesion_rough}).
During pull-off some liquid molecules are 
pulled out of the valleys (or cavities), and form nano-size capillary bridges in the 
asperity regions of the rough substrate (see Fig.~\ref{r_b2_ep40_snap}),
which results in an effective wall-wall interaction which is more long ranged 
than when the wall-liquid 
interaction is weak. For the weak adsorbate interaction case 
(see Fig.~\ref{r_b2_ep5_snap}) the liquid does not wet the surface
of the block (hydrophobic interaction) and no 
capillary bridges form, and for $\theta > 1$ the pull-off stress is 
much smaller than that for the strong adsorbate interaction. 

\begin{figure}
  \includegraphics[width=0.4\textwidth]{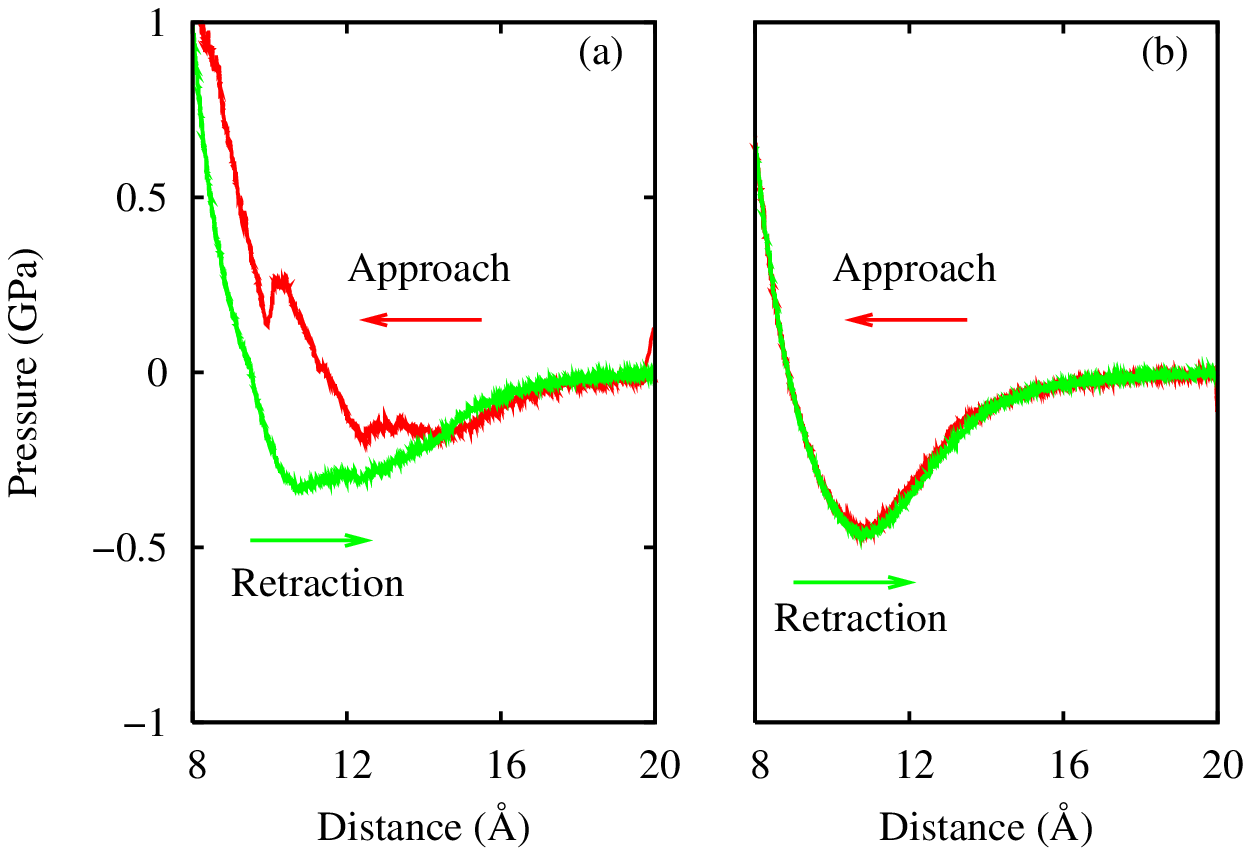} 
  \caption{ \label{r_s3_ep40_ep5_du}
          The average pressure as a function of the distance
          between the block and the rough substrate, with 
          $\theta{\approx}1/4$ monolayer of $C_{8}H_{18}$.
          (a) Strong interaction. (b) Weak interaction
          }
\end{figure}

\begin{figure}
  \includegraphics[width=0.4\textwidth]{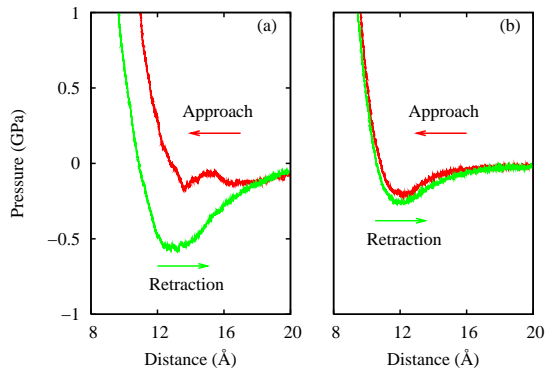} 
  \caption{ \label{r_b2_ep40_ep5_du}
          Average pressure as a function of the distance between
          the block and the rough substrate, with 
          $\theta{\approx}1.3$ monolayer of $C_{8}H_{18}$.
          (a) Strong interaction. (b) Weak interaction
          }
\end{figure}

The force-distance curves for the rough substrate with 
$\theta \approx 0.25$ and $\theta \approx 1.3$, are shown in 
Fig.~\ref{r_s3_ep40_ep5_du} and \ref{r_b2_ep40_ep5_du} 
respectively, both for the strong (a) and weak (b) adsorbate interaction.
Note that for the weak adsorbate interaction there is nearly no hysteresis in the
stress. For the strong adsorbate interaction large hysteresis is observed, and
several abrupt changes in the pressure can be observed during squeezing, which
correspond to the squeeze-out of some lubricant molecules from some
asperity contact regions.

The breaking of the bonds between two macroscopic solids during pull-off
is usually due to interfacial crack propagation, and the macroscopic 
pull-off {\it force} is determined by the {\it energy} 
(per unit of created surface area) $G(v)$ to propagate the crack at the 
velocity $v$, or, equivalently the effective work of 
adhesion $\gamma_{\rm eff}(v)$.
For very small solid objects the bond-breaking may occur more uniformly over 
the contact area
and the standard picture described above no longer holds (see Ref.~\cite{WEAR}).
Thus, the results presented in this study may be directly relevant for 
adhesion involving small (micro- or nano-) sized solid objects,
while the pull-off force for macroscopic objects may be more related to the 
work of adhesion during retraction, which can be obtained from 
pressure-distance plots such as those presented in Fig.~\ref{r_s3_ep40_ep5_du} 
and \ref{r_b2_ep40_ep5_du} but not studied in detail in this paper.

Very recently Huber et al\cite{Huber,thesis} 
have studied the influence of humidity on
the nanoscale adhesion of individual Gecko spatulae to
smooth glass surfaces. The hierarchical gecko
foot structure consists of fiber or hair-like structures, 
and each fiber ends with a
$\sim 100 \ {\rm nm}$ wide plate-like structure which is only a few nanometer
thick at its thinnest place. The plates can easily bend to make contact
even to very rough substrates\cite{Bo_JCP}.
Huber et al observed that the pull-off force on hydrophilic
glass increased
monotonically with increasing humidity. At the highest humidity the water
film was $\sim 0.2 \ {\rm nm}$ thick which correspond
roughly to a water monolayer. This is consistent with our numerical results
for rough
surfaces and hydrophilic materials,
which show that the pull-off force increases with increasing $\theta$ as
$\theta$ increases from 0.2 to monolayer coverage (see Fig. \ref{adhesion_rough}).
As the glass surface used in the experiments by Huber et al was very smooth, 
the dominant surface roughness may be that of the bottom surface of the
plate-like structure of the spatula.
When one spatula is completely submerged under water
the pull-off force decreases by a factor $\sim 6$ as compared to humid condition.

In conclusion, we have studied how molecularly thin liquid layers affect the 
adhesion between the solids both with smooth and rough surfaces.
For strong interaction between the
liquid and walls, the pull-off force exhibits a local {\it maximum} at monolayer
coverage. For the weak interaction, the 
pull-off force decreases continuously with increasing liquid coverage, 
especially when $\theta>1$ for rough substrate.
It is clear that a fundamental understanding of the influence of liquid on 
adhesion at the micro- or nanoscale is central to a large number of biological
and ``high-tech'' engineering applications.

{\bf Acknowledgments}
C.Y. acknowledges the support from the International School for Advanced
Studies (SISSA) and from the Internetional Center of Theoretical
Physics (ICTP) of Triest, Italy.
The contribution of U.T was partly sponsored by
MIUR COFIN No.\ 2003028141-007, MIUR COFIN No.\ 2004028238-002,
MIUR FIRB RBAU017S8 R004, and MIUR FIRB RBAU01LX5H.

\end{document}